\documentclass[]{aa}
\usepackage{natbib}         % pour A&A
\usepackage{graphicx}
\usepackage{longtable}
\usepackage{xcolor}
\bibpunct{(}{)}{;}{a}{}{,} % to follow the A&A style

\def\precision{21}

\newcommand{\ind}[1]{_{\mathrm{#1}}}

\newcommand\Dnu{\Delta\nu}

\newcommand\Teff{T\ind{eff}}
\newcommand\logg{\log g}

\newcommand\HBR{\mathcal{R}}
\newcommand\seuil{\mathcal{E}}
\newcommand\seuildetect{\seuil\ind{det}}
\newcommand\seuilmixte{\seuil\ind{mix}}
\newcommand\seuilDPi{\seuil_{\Delta\Pi}}
\newcommand\seuilrot{\seuil\ind{rot}}

\newcommand\mV{m\ind{V}}

\newcommand{\duree}{{\mathcal{D}}}
\newcommand{\EACF}{{\mathcal{E}}}
\newcommand{\damp}{\kappa}
\newcommand{\EACFd}{{\EACF}_{\damp}}
\newcommand{\rapEACF}{\eta}

\newcommand\numax{\nu\ind{max}}

\newcommand\Hmax{H\ind{max}}
\newcommand\Bmax{B\ind{max}}
\newcommand\Binst{B\ind{noise}}
\newcommand{\mref}{m\ind{ref}}
\newcommand{\Bref}{B\ind{pho}}
\newcommand{\Brefd}{B\ind{add}}\newcommand{\arefd}{a}
\newcommand{\Bsys}{B\ind{sys}}
\newcommand{\seuilM}{\mathcal{E}_M}
\newcommand{\am}{x_M}
\newcommand{\at}{x_{\Teff}}
\newcommand{\fnum}{f}
\newcommand{\adnu}{a_{\Dnu}}
\newcommand{\bdnu}{b_{\Dnu}}
\newcommand{\anum}{a_{\numax}}
\newcommand{\bnum}{b_{\numax}}
\newcommand{\Tred}{T\ind{damp}}
\newcommand{\DTred}{\Delta T\ind{damp}}
\newcommand\Plato{PLATO} %\newcommand\Plato{\emph{Plato}}
\newcommand\Kepler{\emph{Kepler}}

\def\aCenA{$\alpha$\,Cen\,A} \def\aCenB{$\alpha$\,Cen\,B}

\newcommand\suivant{; }
\begin{document}

\title{Seismic performance}
\titlerunning{Seismic performance}
\authorrunning{Mosser et al.}
\author{%
  B. Mosser\inst{1},
  E. Michel\inst{1},
  R. Samadi\inst{1},
  A. Miglio\inst{2,3},
  G.R. Davies\inst{2,3},
  L. Girardi\inst{4},
  MJ. Goupil\inst{1}
  }
\institute{
 LESIA, Observatoire de Paris, PSL Research University, CNRS, Sorbonne Universit\'e, Universit\'e Paris Diderot,  92195 Meudon, France; \texttt{benoit.mosser@observatoiredeparis.psl.eu}
\and
  School of Physics and Astronomy, University of Birmingham, Birmingham, UK
\and
  Stellar Astrophysics Centre, Department of Physics and Astronomy, Aarhus University, Aarhus, Denmark
\and
  INAF – Osservatorio Astronomico di Padova, Padova, Italy
 }

%\date{Submitted to A\&A}

\abstract{Asteroseismology is a unique tool that can be used to study the interior of stars and hence deliver unique information for the studiy of stellar physics, stellar evolution, and Galactic archaeology.}
{We aim to develop a simple model of the information content of asteroseismology and to characterize the ability and precision with which fundamental properties of stars can be estimated for different space missions.
}
{We defined and calibrated metrics of the seismic performance. The metrics, expressed by a seismic index $\EACF$ defined by simple scaling relations, are calculated for an ensemble of stars. We studied the relations between the properties of mission observations, fundamental stellar properties, and the performance index. We also defined thresholds for asteroseismic detection and measurement of different stellar properties}
{We find two regimes of asteroseismic performance: the first where the signal strength is dominated by stellar properties and not by observational noise; and the second where observational properties dominate.  Typically, for evolved stars, stellar properties provide the dominant terms in estimating the information content, while main sequence stars fall in the regime where the observational properties, especially stellar magnitude, dominate.   We estimate scaling relations to predict $\EACF$ with an intrinsic scatter of around 21\%. Incidentally, the metrics allow us to distinguish stars burning either hydrogen or helium.}
{Our predictions will help identify the nature of the cohort of existing and future asteroseismic observations. In addition, the predicted performance for PLATO will help define optimal observing strategies for defined scientific goals.
}


\maketitle

%\voffset = -1.2cm
%________________________________________________________________
\section{Introduction}

The success of the CoRoT\footnote{CoRoT: COnvection, ROtation and planetary Transits, \citep{2006ESASP1306...33B}} and \Kepler\ space-borne missions has opened a new era for stellar physics \citep[e.g.,][]{2008Sci...322..558M,2008A&A...488..705A,2011Sci...332..213C}. Asteroseismology provides unique information on the stars, which is crucial for probing stellar structure and stellar evolution, but also for assessing the physical properties of exoplanets \citep[e.g.,][]{2014A&A...569A..21L} and for Galactic archaeology
\citep[e.g.,][]{2009A&A...503L..21M,2013MNRAS.429..423M,2014MNRAS.445.2758R,2017A&A...597A..30A,2017A&A...600A..70A,2017AN....338..644M}. This success is partly based on the seismic scaling relations that provide relevant estimates of the stellar masses and radii. Seismic radii agree with interferometric results \citep{2012ApJ...760...32H}, even if they are not yet perfectly calibrated, as for instance shown by the comparison with eclipsing binaries \citep{2016ApJ...832..121G}. Solar-like oscillations, excited by turbulent convection in the upper stellar envelope, are observed in late-type stars only. By chance, these stars are numerous and of large interest for
exoplanetology and Galactic archaeology.

Various methods have been set up to detect solar-like oscillations \citep[see, e.g.,][for a comparison of the
methods]{2011MNRAS.415.3539V}. They aim at detecting either the oscillation power in excess with respect to the stellar background, or the regularity of the oscillation pattern. They largely converge on similar results, as recently illustrated by K2 observations \citep{2015PASP..127.1038C, 2017ApJ...835...83S}. In practice, they are used in parallel for the automated detection of oscillation in the wealth of data provided by CoRoT \citep[e.g.,][]{2009Natur.459..398D,2010A&A...517A..22M,2018AN....339..134D}, \Kepler\ \citep[e.g.,][]{2010ApJ...723.1607H,2010A&A...522A...1K,2011Natur.471..608B},
or K2 \citep{2017ApJ...835...83S}: combining the properties of the different pipelines is in fact useful to optimize the seismic yield.

Ensemble asteroseismology has been highly productive. Many scaling relations have been established and used for assessing stellar properties \citep[e.g.,][]{2010A&A...509A..77K,2011ApJ...741..119M}. Recently, they were used for verifying the calibration of the first Gaia data release \citep{2017A&A...599A..50A}. In fact, all  scaling relations are the translation of stellar physics and evolution principles \citep[e.g.,][]{2011A&A...530A.142B}. Scaling relations dealing with the different frequencies or frequency spacings remind us that the surface gravity \citep[or the frequency $\numax$ of maximum oscillation signal,][]{2011A&A...530A.142B} and the mean stellar density \citep[or the large frequency separation $\Dnu$,][]{1917Obs....40..290E} are crucial parameters since they reveal the stellar mass and radius, provided proper effective temperatures are also determined \citep[e.g.,][]{2010A&A...509A..77K,2010A&A...517A..22M,2011ApJ...743..143H}. Scaling relations dealing with the energetic parameters of the stars \citep[e.g.,][]{2011ApJ...732L...5C, 2011ApJ...741..119M,2014A&A...570A..41K}, sustained by theoretical considerations \citep[e.g.,][]{2013A&A...559A..39S,2013A&A...559A..40S} also show that the magnitude of the seismic signal is determined by intrinsic stellar properties. Invaluable information has been derived from ensemble asteroseismology, with recent results dealing with the monitoring of the core rotation in red giants allowing us to characterize the transfer of angular momentum \citep{2018A&A...616A..24G}, the evolution of the lifetime of radial modes \citep{2018A&A...616A..94V}, the thorough analysis of the red giant branch (RGB) bump \citep{2018ApJ...859..156K}, and the study of the chronology of the evolution of the Galactic disk \citep{2018MNRAS.475.5487S}. Altogether, the large success of ensemble asteroseismology motivates the search for a simple index assessing the seismic performance. We define this term as the capability of seismology to derive crucial stellar properties like mass, radius, rotation rate, evolutionary stage, or stellar age.

We perform a comprehensive analysis of the seismic performance in order to answer basic questions related to the detectability of solar-like oscillations, in the same vein as previous work \citep[e.g.,][]{2011ApJ...732...54C,2016ApJ...830..138C}; the precision in measuring stellar parameters like the mass is also addressed. Moreover, we intend to provide an answer without having to simulate oscillation light curves or spectra \citep[e.g.,][]{2018ApJS..239...34B,samadi}. Our approach is therefore less precise than a thorough analysis of all possible sources of noise \citep{samadi} but it is complementary. It is characterized by a simple requirement: the performance has to be expressed analytically in order to address very large set of stars. This choice, initially motivated by the definition of the project SINDICS\footnote{SINDICS: Seismic INDICes Survey}, a long-term, all-sky seismic survey aiming at observing half a million red giants \citep{2015EPJWC.10106045M}, appears to be convenient for all large projects, such as TESS\footnote{TESS:
Transiting Exoplanet Survey Satellite \citep{2015JATIS...1a4003R}.} or \Plato\footnote{PLATO:
PLAnetary Transits and Oscillations of stars \citep{2014ExA....38..249R}.}.

Section \ref{formalisme} presents the theoretical background of our work and explains how the seismic signal is analyzed with the envelope autocorrelation function \cite[EACF, from][hereafter MA09]{2009A&A...508..877M}. Previous observations are used to calibrate the EACF signal and to define an analytical proxy of the seismic performance index. In Sect. \ref{calibration}, observations and theoretical considerations are used to define different thresholds corresponding to different information. The performance index and these thresholds are used together to define the seismic performance.  In Sect. \ref{simulation}, we then depict the simulation tool able to predict this performance. Since the calibration effort has revealed a dependence on the evolutionary stage of evolved red giants, we investigate in Sect. \ref{discussion} a new tool for discriminating between stars before or after the onset of helium burning. We also explore the seismic performance to be reached in open clusters and examine a synthetic field that could be observed by the space mission \Plato\ \citep{2014ExA....38..249R}. Section \ref{conclusion} is devoted to the conclusion.

%---------------------------------------------------------------------------
\section{Method and new calibration\label{formalisme}}

In this work, we define the detection of solar-like oscillations as the measurement of the large separation $\Dnu$, which identifies the frequency spacing between radial modes. This measurement ensures the measurement of $\numax$, which measures the location of the maximum oscillation power, hence the derivation of stellar masses and radii from seismic scaling relations \citep[e.g.,][]{2010A&A...509A..77K}. Results are expressed as a function of $\numax$, since this variable is related to the spectrometric observables $\logg$ and $\Teff$ from the dependence on the acoustic cutoff frequency $\nu\ind{c}$ \citep{1991ApJ...368..599B}:
\begin{equation}\label{eqt-numax-logg}
    \log {\numax \over \numax{}_{\odot}}
    =
     \log {\nu\ind{c} \over {\nu\ind{c}}_{\odot}}
    =
    \log {g \over g_\odot} - {1\over 2} \log {\Teff \over \Teff{}_{\odot}}
    .
\end{equation}
The explanation of this relation by \cite{2011A&A...530A.142B} introduces the Mach number of the turbulent convection.

\subsection{Autocorrelation of the seismic time series: Signal and noise}

The relevance of searching the seismic signal in the autocorrelation of the time series has benefitted from a thorough theoretical analysis \citep{2006MNRAS.369.1491R,2009A&A...506..435R}. In fact, any signature in a seismic time series is repeated after a delay corresponding to twice the travel time of the seismic wave across the stellar diameter, equal to $2/\Dnu$. This delay is easily observed in the autocorrelation of the seismic time series. In practice, this autocorrelation is performed as the Fourier spectrum of the filtered Fourier transform of the time series.

The noise properties of the detection based on the EACF were explored and assessed by MA09. The noise level in the EACF signal is directly extracted from the noise in the time series and the properties of the filter. The mean noise level in the autocorrelation is derived from the fact that the noise statistics is a $\chi^2$ with two degrees of freedom. The EACF signal is then defined and calibrated with Eqs.~2, 5, and 6 of MA09. With these definitions, a pure noise yields a mean value of the autocorrelation signal equal to unity and a  positive detection can be derived from a statistical test based on the null hypothesis.

\subsection{Envelope autocorrelation function}

The earlier paper MA09 found that the best performance of the EACF is reached when the filter used for selecting the oscillation pattern exactly matches the oscillation excess power. The authors derived that the EACF signal $\EACF$, which is a dimensionless number, varies as power laws of $\numax$, of the height-to-background ratio $\HBR$, and of the observation duration $\duree$,
\begin{equation}\label{eqt-EACF-1}
    \EACF \propto \HBR^\alpha \; \numax^\beta\; \duree^\gamma
    ,
\end{equation}
where $\numax$ is expressed in $\mu$Hz and $\duree$ in months.
MA09 calibrated the optimum EACF with six stars, observed by CoRoT during five months at most: a red giant, a
subgiant, two F-type, and two G-type main sequence stars. They inferred $\alpha=1.5$, $\beta=0.9$, and
$\gamma = 1$.

Each term in Eq.~\ref{eqt-EACF-1} plays a specific role:\\
- The relation between the $\EACF$ factor and the observation duration $\duree$ results from the definition of $\EACF$. Regardless of $\duree$, the calibration of $\EACF$ ensures that the mean noise level is unity. Increasing the observation length provides a thinner frequency resolution $\propto\ 1 / \duree$, then a linear increase of the number of frequency bins constructing the signal, hence a linear increase of any relevant signal. The linear dependence of $\EACF$
on $\duree$ was also addressed by \cite{2012A&A...544A..90H}. So, $\gamma = 1$ by definition.

- The term varying as a power of $\numax$ in Eq.~\ref{eqt-EACF-1}
comes from the frequency range where the oscillation power excess is observed. This width accounts for the contribution of the number of observable modes, so one expects the signal $\EACF$ to vary linearly with it. The variation of $\EACF$ in $\numax^{0.9}$ reported by MA09 is in agreement with the fact that the full width at half maximum (FWHM) of the oscillation power varies as $\numax^{0.9}$ \citep{2010A&A...517A..22M}.

\begin{figure}
 \includegraphics[width=9.0cm]{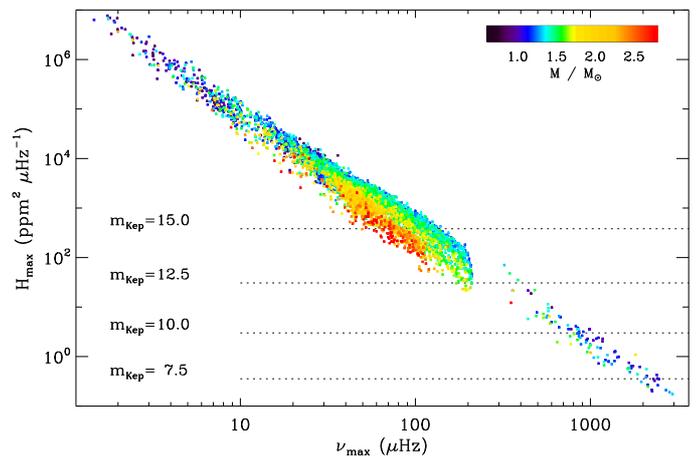}
 \caption{Variation of $\Hmax$ as a function of $\numax$, for stars observed with \Kepler. The color code indicates the seismic estimate of the mass. The instrumental and  photon noise power spectrum density is also indicated by dotted lines for various stellar magnitudes.
 }\label{fig-Hmax}
\end{figure}

- The height-to-background ratio $\HBR$ includes different contributions: the seismic signal, the stellar background, the stellar photon shot noise, and the instrumental noise:
\begin{equation}\label{eqt-hbr}
    \HBR = {\Hmax \over \Bmax + \Binst}
    ,
\end{equation}
where $\Hmax$ is the average oscillation height defined at $\numax$, $\Bmax$ is the background contribution that represents the different scales of granulation, and $\Binst$ includes the photon shot noise and the instrumental noisy contributions. The terms $\Hmax$ and $\Bmax$ are derived from a smoothed spectrum where two components are identified, the background and the oscillation excess power \citep[e.g.,][]{2011ApJ...741..119M}, following the recipe of \cite{2012A&A...537A..30M}. The variations of these terms along stellar evolution  have already benefitted from numerous studies
\citep[e.g.,][]{2011ApJ...743..143H,2012ApJ...760...32H,2014A&A...570A..41K,2018AN....339..134D} and
observations have shown that the ratio $\Hmax / \Bmax$ is largely independent of the evolutionary stage  \citep{2011ApJ...743..143H,2012A&A...537A..30M}. Here we use
\begin{eqnarray}
% \nonumber to remove numbering (before each equation)
  \Hmax &=& 8.0\ 10^6\ \numax^{-2.29},  \label{eqt-Hmax} \\
  \Bmax &=& \Hmax / 3.5,  \label{eqt-HBR-constant}
\end{eqnarray}
where $\numax$ is expressed in $\mu$Hz. Here, $\Hmax$ and $\Bmax$ are measured in a double-sided spectrum, in order to ensure that they can be directly compared to the instrumental noise $\Binst$ estimated  in the time series. Different methods have shown that the scaling relation of $\Bmax$ is a power law of $\numax$ with an exponent close to the one for $\Hmax$ \citep{2012A&A...537A..30M}. This exponent is in fact impacted upon by the contribution of $\Binst$. After the correction of this term, the difference between the exponents of the power laws of $\Hmax$ and $\Bmax$ is smaller than the uncertainties, so that we consider that $\Hmax$ and $\Bmax$ show similar variations, as expressed by Eq. \ref{eqt-HBR-constant}. This equation explains why the mass dependence in the relation $\Hmax (\numax)$ evidenced in previous work \citep[e.g.,][]{2011ApJ...743..143H,2012A&A...537A..30M} and shown in Fig.~\ref{fig-Hmax} can be neglected in the scaling relation (Eq.~\ref{eqt-Hmax}). It plays a minor role except for helium burning stars since it is corrected by a similar mass dependence in $\Bmax$.

When the non-seismic contribution is dominated by the stellar background $\Bmax$, the contribution of $\HBR$ in Eq.~\ref{eqt-EACF-1} is nearly frozen. When the photon or instrumental noise prevails, $\HBR$  is dominated by the rapid variation of $\EACF$ with stellar evolution. In MA09, the exponent of $\HBR$  equal to 1.5 was determined in a pure phenomenological manner. However, not enough stars were used to provide a robust calibration.

\begin{figure*}
\sidecaption
\includegraphics[width=14.8cm]{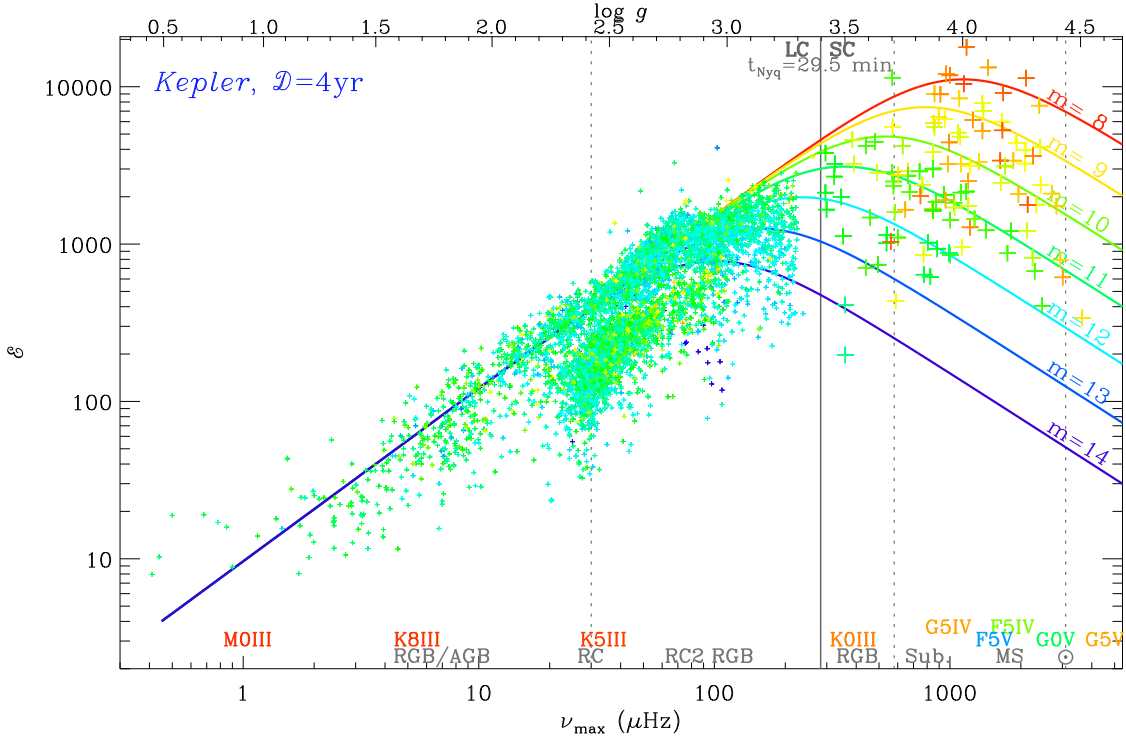}
% calib_kepler de figures_eacf.pro
  \caption{EACF level for \Kepler\ data, homogenized to a uniform observation duration of 48 months, expressed as a function of $\numax$. The $x$-axis also shows the corresponding $\log g$. The color designates the stellar magnitude, as indicated by the theoretical level overplotted on the diagram. The three vertical dotted lines indicate, from low to high frequency, the mean location of the red clump, the transition between subgiants and red giants, and the Sun. The vertical solid line indicates the Nyquist frequency of the long-cadence mode of \Kepler.  Different stellar types and classes are shown parallel to the x-axis, with an independent color code.}\label{fig-calibration-kepler}
\end{figure*}

\subsection{Different noise contributions}

In order to refine the calibration of Eq.~\ref{eqt-EACF-1}, it is necessary to model the term $\Binst$ for each mission. We therefore used the information provided for CoRoT by \cite{2009A&A...506..411A} and for \Kepler\ by \cite{2010ApJ...713L.120J}. We also used the photometric noise models of \Plato\ \cite[Fig. 14 of][]{2014ExA....38..249R} and of TESS \cite[Fig. 3 of][]{2016ApJ...830..138C}. The noise level is typically determined by a simplified conservative model with three components, all expressed in terms of power spectrum densities:
\begin{equation}\label{eqt-Binst-bruits}
    \Binst (m)
    =
    \Bsys + \Bref\ 10^{0.4(m-\mref)} + \Brefd\ 10^{\arefd (m-\mref)}
    .
\end{equation}
The term $\Bsys$ represents the minimum instrumental noise and stands for systematic noise and/or the saturation of the pixel well when bright stars are observed. It shows a frequency-dependent spectrum, which rapidly increases below a few microhertz only \citep{samadi}, not considered here since it affects evolved red giants only. The term  $\Bref$ depends on the photon shot noise of the star. For dim stars, an extra contribution $\Brefd$ accounts for many possible noisy contributions like the sky background, the jitter, or the readout noise \citep{2010ApJ...713L..87J}. For convenience, we chose a reference $\mref$ that was equal for all missions. The contribution of $\Bsys$ is important for bright stars only, and the contribution of $\Brefd$ for dim stars only, so that in most cases corresponding to the core mission, the photon shot noise dominates.

As $\Binst$ is expressed as a power spectral density, it does not vary with $\duree$. The values of $\Bsys^{1/2}$, $\Bref^{1/2}$, $\Brefd^{1/2}$, and $\arefd$ are given in Table~\ref{tab-perf} for the reference magnitude $\mref$.

\begin{table}
\caption{Instrumental noise (with $\mref=8$).}\label{tab-perf}
\small
\begin{tabular}{ccccc}
  \hline
  Project &$\Bsys^{1/2}$ &$\Bref^{1/2}$ &$\Brefd^{1/2}$& $\arefd$  \\
          &\tiny (ppm\,hr$^{1/2}$)&\tiny{(ppm\,hr$^{1/2}$)}&\tiny (ppm\,hr$^{1/2}$)&  \\
  \hline
  CoRoT$^{(a)}$   & 3.2  & 15.5&     &      \\
  \Kepler$^{(b)}$ & 2.8  &  8.0& 0.4 & 0.7 \\
%  \Plato$^{(c)}$  & 9.4  & 9.8& 2.1 & 0.66 \\
  \Plato$^{(c)}$  &  9  & 9& 2 & 0.66 \\
  TESS$^{(d)}$    &0 -- 60& 60  & 10  & 0.68 \\
  \hline
\end{tabular}

\small

$(a)$ \cite{2008A&A...488..705A} and \cite{2009A&A...506..411A}; as only bright stars were considered, we did not model the additional component.\\
$(b)$ \cite{2010ApJ...713L.120J}; the systematic noise is derived from the observation of the bright twin systems 16 Cyg A and B by \cite{2012ApJ...748L..10M}.
\\
$(c)$ \cite{2014ExA....38..249R}, updated to the new configuration of \Plato\ with 24 telescopes and following the requirements of the mission expressed in the PLATO Definition Study Report. \\
$(d)$ \cite{2016ApJ...830..138C}; $\Bsys^{1/2}=60$ ppm hr$^{1/2}$ is presented as the worst case; we use $\Bsys^{1/2}=30$ ppm hr$^{1/2}$ in our simulations.
\end{table}

\subsection{New calibration of the EACF\label{newcalibration}}

In order to update the calibration of $\EACF$, we took advantage of the large sets of stars observed by CoRoT and \Kepler, either in the short-cadence mode \citep[e.g.,][]{2008Sci...322..558M,2011Sci...332..213C,2014ApJS..210....1C}, or in the long-cadence mode \citep[e.g.,][]{2010A&A...517A..22M,2013ApJ...765L..41S}, all reprocessed with the EACF in an identical way. We considered the data available on the KASOC\footnote{KASOC: Kepler Asteroseismic Science Operations Center} database \citep{2014MNRAS.445.2698H}, avoiding stars close to the red edge of the instability strip since their oscillations are significantly damped \citep{2011ApJ...732...54C}, and excluding core-helium burning stars. This is possible since more than 6000 evolved stars have a firm measurement of their evolutionary stage derived from the analysis of their period spacings \citep{2015A&A...584A..50M,2016A&A...588A..87V} or from other methods \citep{2017MNRAS.466.3344E}. We fit the exponents $\alpha$ and $\beta$ of Eq.~\ref{eqt-EACF-1} and got the new calibration
\begin{equation}\label{eqt-EACF-norm}
    \EACF = 0.065\; \HBR^{0.9} \; \numax^{1.1}\; \duree
    ,
\end{equation}
where $\numax$ is expressed in $\mu$Hz and $\duree$ in months. We note that the exponents are very close to
but do not match the first values derived by MA09. A different exponent for $\HBR$, 0.9 instead of 1.5, is not surprising, since the six CoRoT stars had only low values of $\HBR$. The small change of $\alpha$ is then counterbalanced by a change of $\beta$, which, however, remains close to the value 0.9 previously found.

The calibration of $\EACF$ with \Kepler\ data is shown in Fig.~\ref{fig-calibration-kepler}
where, in order to make the comparison possible, we neutralized the duration dependence by considering a uniform observation duration and corrected $\EACF$ accordingly. All stars follow the calibration reported by Eq.~\ref{eqt-EACF-norm} except in three cases:

\begin{enumerate}
  \item Very high-luminosity red giants with $\numax\le 1.5\,\mu$Hz show $\EACF$ in excess with the fit. This is in line with the fact that these semi-regular variables do not oscillate purely in the solar-like mode, but evolve toward the
Mira stage \citep{2013A&A...559A.137M}. Therefore, they require a specific treatment beyond the scope of this work. Moreover, their observation with future space-borne missions requires a long duration (at least as long as the four-year length provided by \Kepler) and is, unfortunately, unlikely.
   \item Red-clump (RC) stars, with $\numax$ in the frequency range [25 - 125\,$\mu$Hz], behave differently and require special care. In fact, the EACF of clump stars is lower compared to RGB stars, so that the fit provided by Eq.~\ref{eqt-EACF-norm} has to be modified by a factor 0.41:
\begin{equation}\label{eqt-EACF-norm-RC}
    \EACF\ind{RC} = 0.027\; \HBR^{0.9} \; \numax^{1.1}\; \duree
    .
\end{equation}
This specific behavior of stars in the red clump or on the asymptotic giant branch is discussed in Sect. \ref{evolution}.\\
    \item Stars close to the instability strip show a damped signal, as reported by \cite{2011ApJ...732...54C} and \cite{2011ApJ...743..143H}. The expression for the damping of the mode amplitudes proposed by their work can be directly used for main sequence stars and subgiants, but not for red giants. We therefore have to consider the damping factor
\begin{equation}\label{eqt-damping}
  \damp = 1 - \exp\left( {\Teff - \Tred \over \DTred} \right)
\end{equation}
with adapted parameters
\begin{eqnarray}
% \nonumber to remove numbering (before each equation)
  \Tred  &=& {\Tred}_\odot\, \left({1\over M} {\numax \over {\numax}_\odot} {{{\Teff}_\odot}^{3.5} \over {\Teff}^{3.5}} \right)^{0.093}, \label{Tdamp} \\
  \DTred &=&  {\DTred}_\odot\, \left( {\numax \over  {\numax}_\odot} \right)^{0.16}. \label{DTdamp}
\end{eqnarray}
The reference temperatures  ${\Tred}_\odot=8907$\,K and ${\DTred}_\odot=1550$\,K  are defined in \cite{2011ApJ...732...54C}. Equation~\ref{Tdamp} represents the power law varying in $L^{0.093}$ considered in their work; here, we have replaced the luminosity by its seismic proxy $M\,\numax^{-1}\, \Teff^{3.5}$. The power law expressed by Eq.~\ref{DTdamp} fits with our measurements.
\end{enumerate}

For all evolutionary stages, the ratio $x$ between the observed and synthetic values of the EACF is close to a normal distribution with a standard deviation of {0.\precision}. Possible physical reasons explaining this spread are discussed below in Sect. \ref{spread-raisons}. %Anyway, the spread in $x$ directly translates into an estimate of the precision of the fit; this value is therefore typical of the uncertainty level of the simulation of the seismic performance.

\begin{figure*}
 \sidecaption
 \includegraphics[width=12cm]{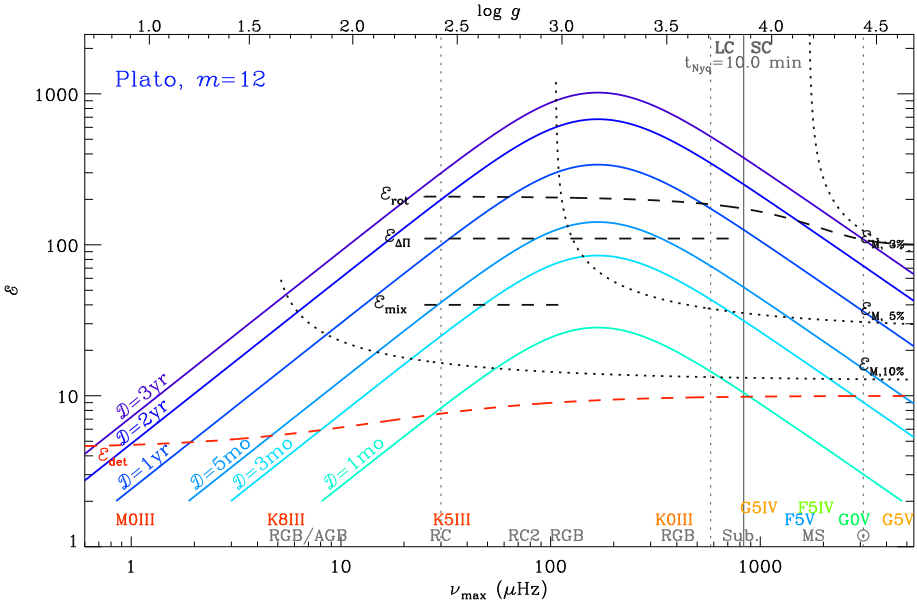}
 % simu_duree
 \caption{Variation with $\numax$ of the performance index $\EACF$ for a $\mV=12$ star observed by \Plato\ for different durations.  We use the same style as in Fig.~\ref{fig-calibration-kepler}, except that the color code of the curves represents the observation duration. The different thresholds corresponding to the different cases established by Eqs.~\ref{eqt-seuil-detect} to \ref{eqt-seuil-rot} are plotted with dashed lines; the thresholds defining the precision one can expect from the seismic relations (Eq.~\ref{eqt-seuil-dlogm}) are plotted with dotted lines.
 }\label{fig-simu-plato-duree}
\end{figure*}
\begin{figure*}
 \sidecaption
 \includegraphics[width=12cm]{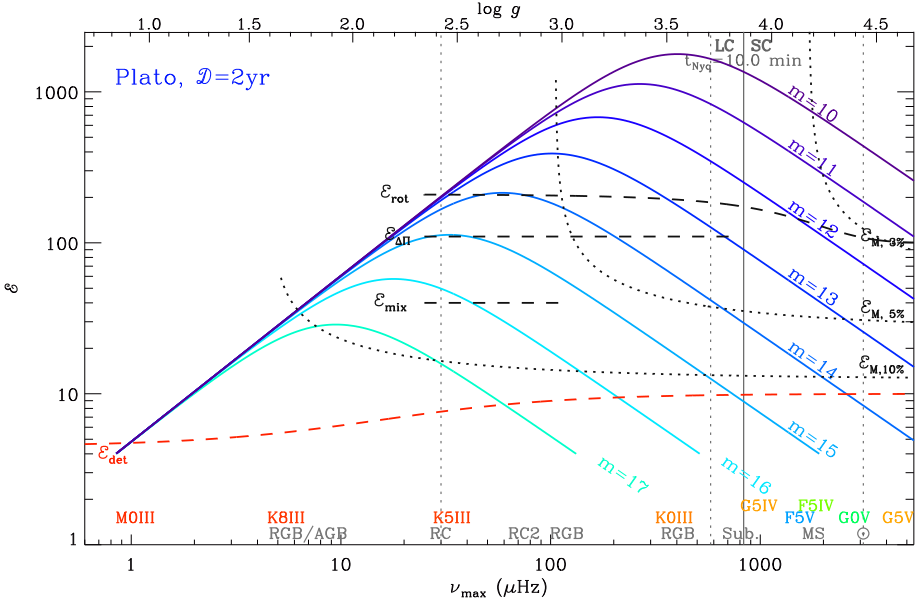}
 \caption{Variation with $\numax$ of the performance index $\EACF$ for a two-year observation run of \Plato, for  different stellar  magnitudes. We use the same style as Figs.~\ref{fig-calibration-kepler} and \ref{fig-simu-plato-duree}, except that the color code of the curves represents the stellar magnitude.
 }\label{fig-simu-plato-mag}
\end{figure*}

%---------------------------------------------------------------------------
\section{Assessing the performance\label{calibration}}

In the previous section, we  calibrated the EACF signal $\EACF$. This calibration makes it possible to use $\EACF$  as information that is no longer related to the EACF, but to stellar and instrumental properties, and reveals the  intrinsic properties of $\EACF$ (Fig. \ref{fig-calibration-kepler}). From this stage on, we leave the link between $\EACF$ and the EACF signal and consider $\EACF$ as a measurement of the seismic signal. We use it as a performance index that we have to compare with different thresholds associated with different conditions in order to assess the seismic performance.

\subsection{Positive seismic detection}

The threshold for a seismic detection is derived from a null hypothesis (H0) test: MA09 showed that the H0 hypothesis is rejected at the {1\,\%}-level when $\EACF$ is larger than 8.0. The rejection at the {10\,\%}-level is ensured for $\EACF \ge 5.7$. These values are a priori valid for any values of $\Dnu$, so at any evolutionary stage. However, the detection of solar-like oscillation is obtained at lower $\EACF$ for the most evolved stars \citep{2013A&A...559A.137M}, whereas a higher $\EACF$ of 10 is most suitable for main sequence stars. We therefore redefined the threshold of positive detection accordingly:
\begin{equation}\label{eqt-seuil-detect}
    \seuildetect =
    \left\{
    \begin{array}{rl}
        5.5  & \mbox{if }  \numax \le 10\, \mu\mbox{Hz}, \\
        8    & \mbox{for } 20 \le  \numax \le 500\, \mu\mbox{Hz}, \\
        10   & \mbox{if }  \numax \ge 800\, \mu\mbox{Hz},
    \end{array}
    \right.
\end{equation}
with smooth transitions between the different frequency ranges.
The coefficient $\EACF$ can also be compared to different thresholds above which the seismic signal makes it possible to identify the fine structure of the oscillation spectrum, and thus to derive useful information in terms of stellar structure.

\subsection{Threshold for determining the core properties of evolved stars}

Unique information is provided by the mixed modes present in evolved stars with a radiative core.
Their detection allows us to disentangle core-helium-burning from hydrogen-shell-burning stars \citep{2011Natur.471..608B,2011A&A...532A..86M}. Empirically, we defined the threshold value
\begin{equation}\label{eqt-seuil-mixte}
    \seuilmixte = 40
    ,
\end{equation}
above which their detection is possible, and hence the discrimination of the evolutionary stage in the frequency range [3, 120\,$\mu$Hz]. Having a nearly flat threshold comes from the expression of $\EACF$. In fact, resolving mixed modes requires the ability to identify small frequency spacings with respect to $\numax$; such a comparison is driven by the frequency resolution, proportional to $\duree^{-1}$. Owing to the fact that the ratio $\Hmax/ \Binst$ is uniform for most stars where this study is relevant, and that the dependence in $\numax$ has an exponent close to 1 (Fig.~\ref{fig-calibration-kepler}), a comparison between a frequency proportional to $1/\duree$ and $\numax$ translates into a simple condition on $\EACF$. In the red clump, the value $\seuilmixte$ is typically reached after four months, according to Eq.~\ref{eqt-EACF-norm}: CoRoT observations lasting five months were indeed enough for identifying RGB stars \citep{2011A&A...532A..86M}.

The measurement of the period spacing associated to the gravity component of mixed modes is crucial for investigating the core properties of red giants \citep[e.g.,][]{2013ApJ...766..118M,2016MNRAS.457L..59L,2017MNRAS.469.4718B}. We used previous observations to calibrate the corresponding threshold $\seuilDPi$. A long enough measurement and a higher $\EACF$ value are required compared to the simple identification of the evolutionary stage. We found that the threshold can be considered as uniform in the frequency range $\numax < 800\,\mu$Hz where mixed modes are observed in subgiants and red giants:
\begin{equation}\label{eqt-seuil-DPi}
    \seuilDPi  =
    \left\{
    \begin{array}{rl}
        110 & \mbox{on the  RGB} \\
         80 & \mbox{in the red clump}
    \end{array}
    \right.
.\end{equation}
The justification of this uniform value is more complicated than for $\seuilmixte$. In fact, the measurement of narrower period spacings when the stars evolve on the RGB is more demanding than for the early RGB, but this difficulty is compensated for by a larger number of mixed modes. The traduction of the thresholds $\seuilDPi$ in terms of the observations' duration gives typically an observation running eight months for a clump star or 11 months for a RGB star.

\begin{table*}[t]
   \caption{Summary of the seismic simulation.}\label{tab-simu}
  \small
  \begin{tabular}{llllc}
%    \hline
%   & Independent parameters         &                   & Stellar  & Instrumental\\
%   &                   &                   & regime   & regime \\
%  \hline
%  & Stellar parameters & $M$, $R$, $\Teff$ & $\times$ & $\times$  \\
%  & Stellar magnitude  &     $m$           &          & $\times$  \\
%  & Noises             & $\Bsys,\Bref,\Brefd$ &        & $\times$  \\
%  & Observation length & $\duree$          &          &           \\
     \hline
       & Steps & Input & Output & Equations\\
     \hline
     $[0]$   & $\numax$ from stellar parameters &  $M$, $R$, $\Teff$, or $\logg$ &  $\numax$ & (\ref{eqt-numax-logg}) \\
     $[1.1]$ & Oscillation signal and granulation background& $\numax$ & $\Hmax$, $\Bmax$ & (\ref{eqt-Hmax}),  (\ref{eqt-HBR-constant}) \\
     $[1.2]$ & Instrumental noise & $m$, $\Bsys$, $\Bref$, $\Brefd$  & $\Binst$ & (\ref{eqt-Binst-bruits}) + Table \ref{tab-perf} \\
     $[1.3]$ & Height-to-background ratio           & $\Hmax$, $\Bmax$, $\Binst$ & $\HBR$ & (\ref{eqt-hbr}) \\
     $[2.1$H] & Raw performance index ($\le$ RGB stage) &  $\HBR$, $\numax$, $\duree$ & $\EACF$ & (\ref{eqt-EACF-norm})\\
     $[2.1$He]& Raw performance index ($\ge$ RC stage) &  $\HBR$, $\numax$, $\duree$ & $\EACF$ & (\ref{eqt-EACF-norm-RC})\\
     $[2.2]$ & Damping close to the instability strip& $M$, $\Teff$, $\numax$ &  $\damp$ &  (\ref{eqt-damping})\\
     $[2.3]$ & Seismic index  & $\EACF$, $\damp$ & $\EACFd = \damp\ \EACF$ & \\
     $[3.1]$ & Detection threshold& $\numax$ & $\seuildetect$  & (\ref{eqt-seuil-detect}) \\
     $[3.2]$ & Thresholds: evolutionary stage, rotation & $\numax$ & $\seuilmixte$, $\seuilDPi$, $\seuilrot$ & (\ref{eqt-seuil-mixte}) $\to$ (\ref{eqt-seuil-rot})\\
     $[3.3]$ & Thresholds: relative mass uncertainty & $\numax$, $\am$ & $\seuilM$  & (\ref{eqt-seuil-dlogm})\\
     $[4.1]$ & Positive seismic detection & $\EACFd$, $\seuildetect$ & \multicolumn{2}{l}{$\EACFd \ge \seuildetect$} \\
     $[4.2]$ & Seismic performance & $\EACFd$, $\seuilmixte$, $\seuilDPi$, $\seuilrot$, $\seuilM$ &
      \multicolumn{2}{l}{$\EACFd \ge \seuilmixte$, or $\EACFd \ge...$}  \\
     \hline
   \end{tabular}

   The independent variables are the stellar parameters $M$, $R$, $\Teff$, the stellar magnitude $m$, the instrumental noise contributions $\Bsys$, $\Bref$ and $\Brefd$, and the observation length $\duree$ (see Table \ref{tab-regimes}).
\end{table*}

\subsection{Threshold for rotation}

Measuring the stellar rotation rate is of prime importance, owing to its role in stellar structure and evolution. Seismic observations allows us to probe the envelope rotation in main sequence stars \citep{2014A&A...566A..20A} and the mean core rotation in RGB stars \citep{2012Natur.481...55B,2012A&A...540A.143M,2017A&A...605A.111C,2018A&A...616A..24G}. The observation of mixed modes provides both values, in subgiants and low RGB stars \citep{2014A&A...564A..27D}.
As above, we used previous observations to calibrate the thresholds
\begin{equation}\label{eqt-seuil-rot}
    \seuilrot =
    \left\{
    \begin{array}{rl}
        250  & \mbox{for }  70 \le \numax \le 500\, \mu\mbox{Hz on the RGB}, \\
        120  & \mbox{for }  30 \le \numax \le 120\, \mu\mbox{Hz in the red clump}, \\
        250  & \mbox{for }  500 \le \numax \le 800\, \mu\mbox{Hz for subgiants}, \\
        100  & \mbox{if }  \numax \ge 800\, \mu\mbox{Hz},
    \end{array}
    \right.
\end{equation}
with a smooth transition between the different frequency ranges. The threshold concerning rotation is valid everywhere, except for evolved stars after the red clump for which no seismic observation of rotation has been reported due to the lack of mixed modes \citep{2018A&A...618A.109M}. Such a threshold is demanding for evolved stars: observing the rotation in the red clump requires an observation longer than two years.

\subsection{Uncertainty on the seismic stellar mass}

A different way to express the seismic performance consists in estimating the precision we can infer for the stellar mass given by the seismic scaling relations. This precision is derived from the relative uncertainties on $\Dnu$ and $\numax$, which were calibrated with \Kepler\ data as a function of $\EACF$ as
\begin{eqnarray}
% \nonumber to remove numbering (before each equation)
  \delta\log \Dnu   &=& \adnu + \bdnu \, \EACF^{-1} \\
  \delta\log \numax &=& \anum + \bnum \, \EACF^{-1}
  .
\end{eqnarray}
The relevance of these uncertainties was also checked by the comparison with other methods \citep{2011MNRAS.415.3539V,2011A&A...525A.131H,2018ApJS..239...32P}.
We noted that the relative precision in $\numax$ is dominated by the term $\anum$,  defined as a fraction $\fnum$ of the large separation $\Dnu$. This assumption relies on the fact that measuring $\numax$ remains method dependent \citep[e.g.,][]{2018ApJS..239...32P}. Moreover, the definition of the power excess still lacks a precise theoretical support and the assumption of its Gaussian shape is known to be inadequate in some cases, as was clearly noted for Procyon \citep{2008ApJ...687.1180A,2010ApJ...713..935B}. As a result, the relative precision in $\numax$ does not decrease when the EACF signal increases. Assuming $\bnum\equiv 0$ as justified above, we consider
\begin{eqnarray}
% \nonumber to remove numbering (before each equation)
  \anum  &=& \fnum \Dnu / \numax \ \hbox{ and } \ \fnum\ =\ 1/6; \\
  \adnu  &=& 7.5\times 10^{-4}; \\
  \bdnu  &=& 0.3.
\end{eqnarray}
The relative mass uncertainties $\am$ in $M$ are then derived from the uncertainties on $\Dnu$ and $\numax$ propagated through the scaling relation $M\propto {\numax}^3 \, {\Dnu}^{-4}\, {\Teff}^{-3/2}$. An uncertainty of 80\,K on $\Teff$ is included, corresponding typically to a relative uncertainty $\at$ of about 1.5\,\%. As $\am$ depends on $\EACF$, it possible to define the threshold $\seuilM(\am)$ on the EACF necessary to obtain a given uncertainty $\am$; it expresses
\begin{equation}\label{eqt-seuil-dlogm}
  \seuilM(\am) = {4\,\bdnu \over \displaystyle{
            \sqrt{\am^2 - \left(3 \,{\fnum\,\Dnu \over \numax} \right)^{2} - \left({3\over2}\,\at\right)^2} - 4\,\anum   }}
  .
\end{equation}
The values of $\seuilM$ are shown in Figs.~\ref{fig-simu-plato-duree} and \ref{fig-simu-plato-mag} for a relative precision $\am$ equal to 3, 5, or 10\,\% and for $\at=1.5$\,\%. A relative mass uncertainty $\am$ typically corresponds to a relative age uncertainty of $3\ \am$ at all evolutionary stages, including red giants. This proxy accounts for the fact that stellar ages before the main-sequence turnoff are expected to scale with $M^3$, and that stellar evolution after the turnoff is much more rapid than before \citep{2013MNRAS.429..423M,2017AN....338..644M}.

It is worth mentioning that the uncertainties we provide are representative of the use of the raw seismic scaling relations based on global seismic parameters. Following various work  \citep[e.g.,][]{2010A&A...522A...1K,2014ApJS..214...27M}, we consider that using grid-based modeling typically doubles the precision we can infer on stellar radii and masses . In the figures reporting the simulation, we consider that the threshold for $\am=5$\,\% obtained with the scaling relations (Eq.~\ref{eqt-seuil-dlogm}) is representative of a precision of 3\,\% on the stellar mass after grid-based modeling. Therefore, this threshold defines the domain where stellar ages can be estimated at the 10\,\%-level from the determination of individual oscillation frequencies.

%---------------------------------------------------------------------------

\section{Simulation\label{simulation}}

The calibration of $\EACF$ allows us to simulate the seismic performance of any stellar field of view. We consider here the observation of a star showing solar-like oscillations and observed with precise photometry, as done by the space missions CoRoT, \Kepler, \Plato, or TESS.

\begin{table}
\caption{Stellar and instrumental regimes.}\label{tab-regimes}
\small
\begin{tabular}{llcc}
  \hline
  Parameters         &                   & Stellar  & Instrumental\\
                     &                   & regime   & regime \\
  \hline
  Stellar parameters & $M$, $R$, $\Teff$ & $\times$ & $\times$  \\
  Stellar magnitude  &     $m$           &          & $\times$  \\
  Noises             & $\Bsys,\Bref,\Brefd$ &        & $\times$  \\
  Observation length & $\duree$          &          &           \\
  \hline
\end{tabular}

Contribution of the independent input parameters of the simulation to the different regimes.
\end{table}

\subsection{Step-by-step tutorial}

The step-by-step assessment of the seismic performance is summarized in Table \ref{tab-simu}, which presents the input, output, and equations of each step. The number of independent inputs is in fact limited: the data required for the simulation are stellar parameters (mass, radius, or $\logg$, and effective temperature, or stellar type as exposed in Table \ref{tab-valeurs}), stellar magnitude, instrumental data (Table \ref{tab-perf}), and the observation duration $\duree$. These independent variables are summarized in Table \ref{tab-regimes}.

\begin{description}
  \item{[0]} This preliminary step defines the global seismic parameter $\numax$ as a function of stellar parameters, which can be derived from spectroscopy ($\logg$, $\Teff$) or from modeling ($M$, $R$, $\Teff$). Another way to connect the stellar classes and types to the seismic stellar parameters is provided by Table \ref{tab-valeurs}, where we show stars with their global seismic parameters $\Dnu$ and $\numax$, their effective temperature, $\logg$, and spectral type.
  \item{[1]} Steps 1.1 to 1.3 are used to construct the height-to-background ratio. They can be adapted to any instrument. The method was, however, tested neither for ground-based observations with limited duty cycle, nor for spectrometric observations.
  \item{[2]} The derivation of the seismic index is a simple step of the simulation. One has to pay attention to the different scaling, depending on the evolutionary stage: [2.1H] concerns stars before helium-core burning, whereas [2.1He] concerns stars after the ignition of core helium. The performance index is then $\EACFd = \damp\, \EACF$, with $\damp$ accounting for the mode damping. For simplicity, all figures represent $\EACF$, except the simulation for Plato presented in Sect. \ref{simuplato}. This simplification, used to consider $\numax$ as the single variable instead of $\numax$ and $\Teff$, does not hamper the conclusion of the paper.
  \item{[3]} The definition of the thresholds is an important ingredient of the method. The predicted $\EACF$ and the threshold values are all in line with the EACF method exposed by MA09; the calibration with CoRoT and \Kepler\ data ensures the consistency of the scalings, regardless of the exact recipe used in MA09. However, as stated earlier, the performance index $\EACF$ defined by Eq.~\ref{eqt-EACF-norm} is no longer related to the EACF computation and is now used independently.
  \item{[4]} A positive detection requires $\EACFd \ge \seuildetect$. The seismic performance then depends on the adopted criterion. The seismic estimate of the stellar mass with a relative precision of 10\,\% is ensured for all unevolved stars showing solar-like oscillations (with the caveat that scaling relations are not yet fully calibrated, which adds some systematic uncertainty upon the statistical uncertainties). The detection of rotation is the most demanding criterion.
\end{description}

\subsection{Stellar and instrumental regimes}

The simulation throws light on the origin of the two regimes observed in  \Kepler\ data. At low frequency (Figs. \ref{fig-calibration-kepler}-\ref{fig-simu-plato-mag}), the stellar magnitude plays no role since the oscillation amplitude is much larger than the different noisy variations. In this stellar regime characterized by $\HBR \simeq 3.5$ (Eq.~\ref{eqt-HBR-constant}), the seismic signature increases with increasing $\numax$, despite the decrease of the oscillating amplitude:
\begin{equation}\label{eqt-perf-bf}
    \EACF\ind{LF} = 0.20\; \numax^{1.1}\; \duree
    .
\end{equation}
This behavior derives from the properties of the oscillation spectrum, with a reduced number of oscillation modes when stars evolve on the RGB.

At high frequency, the seismic performance is dominated by the stellar magnitude. In the main sequence, the
decrease of the oscillation signal with increasing $\numax$ (or with decreasing $\Teff$) induces a rapid decrease of the performance. In this instrumental regime, the performance varies as
\begin{equation}\label{eqt-perf-hf}
    \EACF\ind{HF} \propto \numax^{-0.96} \; 10^{0.36 (\mref - m)}\; \duree
    .
\end{equation}
The exponent in $\numax$ varies as $\beta+\alpha  \, h$, where $\alpha$ and $\beta$ are defined by Eq.~\ref{eqt-EACF-norm}, and where $h$ is the exponent of the scaling relation of $\Hmax$ (Eq.~\ref{eqt-Hmax}).

The limit between the two regimes does not depend on $\duree$, but on the stellar magnitude: the brighter the magnitude, the broader the stellar regime. We show different cases in Figs.~\ref{fig-simu-plato-duree} and \ref{fig-simu-plato-mag}, in the condition of \Plato\ observations, with different observation durations and different stellar magnitudes. From these case studies, it is easy to derive the conditions on the stellar magnitude or observation duration to detect oscillation at any threshold.

\begin{figure}
\includegraphics[width=9cm]{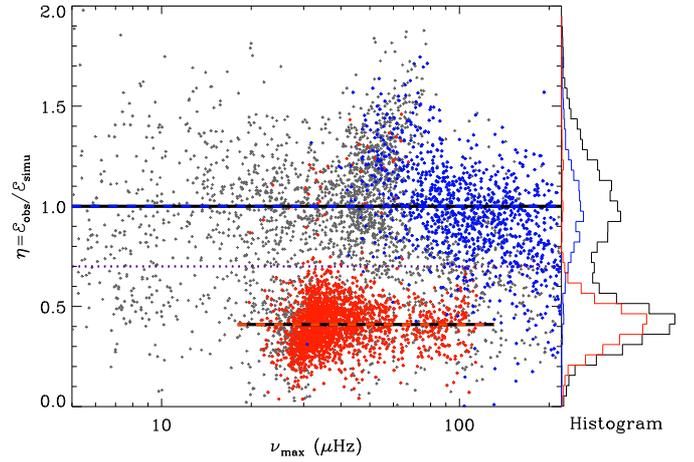}
 \caption{Distribution of the ratio $\rapEACF$ of the observed EACF
compared to the predicted performance (Eq.~\ref{eqt-EACF-norm}), for evolved stars.
 Blue (red) symbols indicate RGB (red clump) stars, as identified by \cite{2016A&A...588A..87V}. The horizontal dashed lines indicate the mean value of the ratio for each evolutionary stage; the dotted line indicates the separation between RGB and RC stars. The right panel shows the bimodal distribution of $\rapEACF$. }\label{fig-spread}
\end{figure}

\begin{figure}
%figure_eacf
\includegraphics[width=9cm]{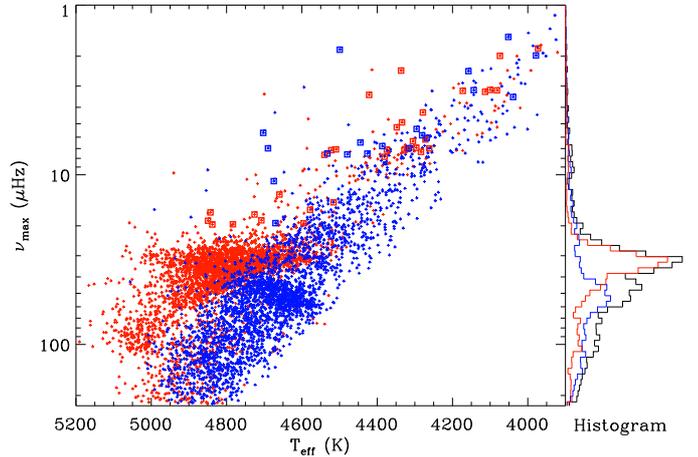}
 \caption{Seismic diagram of the \Kepler\ data, with $\Teff$ derived from the APOKASC survey and $1/\numax$ used as a proxy for the luminosity. Stars identified on the RGB from the $\rapEACF$ parameter are shown in blue; stars identified as clump stars or on the AGB are plotted in red. Evolved low-mass stars are indicated with squares. The right panel shows the histogram of the stellar distribution as a function of $\numax$.}
\label{fig-HR}
\end{figure}

\begin{figure*}
 \sidecaption
 \includegraphics[width=12cm]{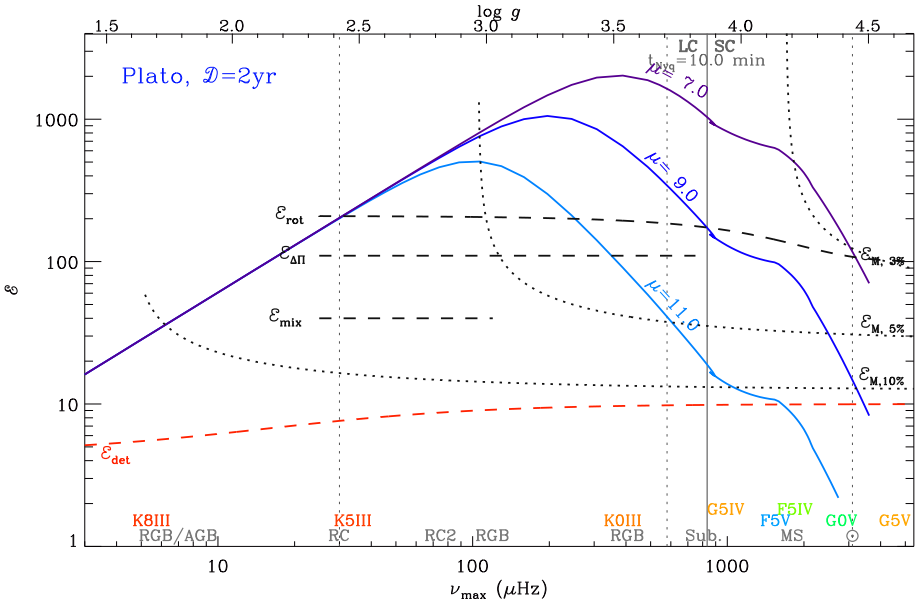}
 \caption{Variation with $\numax$ of the performance index $\EACF$
 for a two-year observation run of \Plato, for open clusters with different distance modules $\mu$. We use the same indications as in Figs.~\ref{fig-calibration-kepler} and \ref{fig-simu-plato-duree}.
 }\label{fig-simu-plato-cluster}
\end{figure*}

\begin{figure*}
 \sidecaption
 \includegraphics[width=12cm]{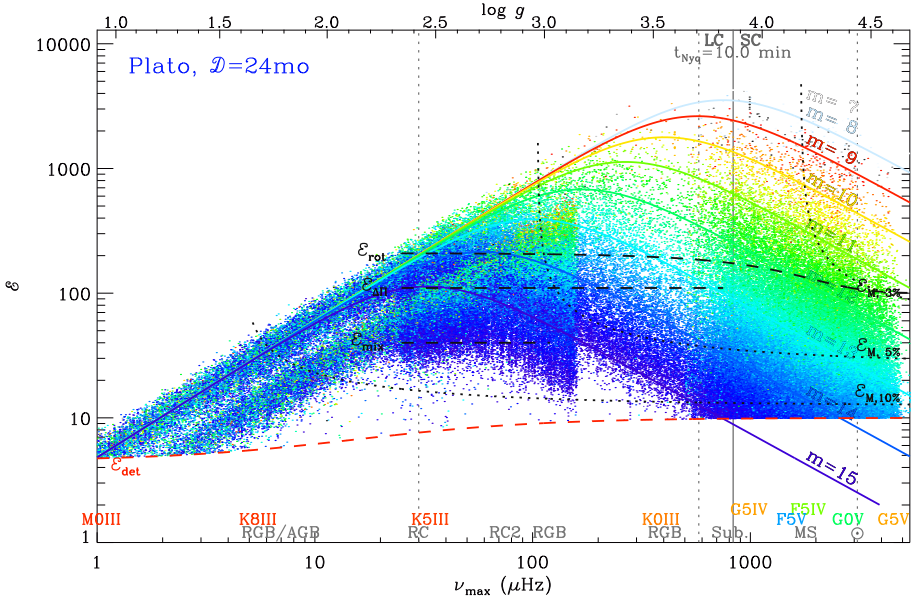}
 \caption{Simulation of a \Plato\ field of view with 1.5 10$^5$ stars brighter than $m\ind{V}=15$, among which 10$^5$ show detectable solar-like oscillations. We use the same indications as in Figs.~\ref{fig-calibration-kepler} and \ref{fig-simu-plato-duree}.}
 \label{fig-simu-champ-plato}
\end{figure*}

\begin{figure*}
 \sidecaption
 \includegraphics[width=12cm]{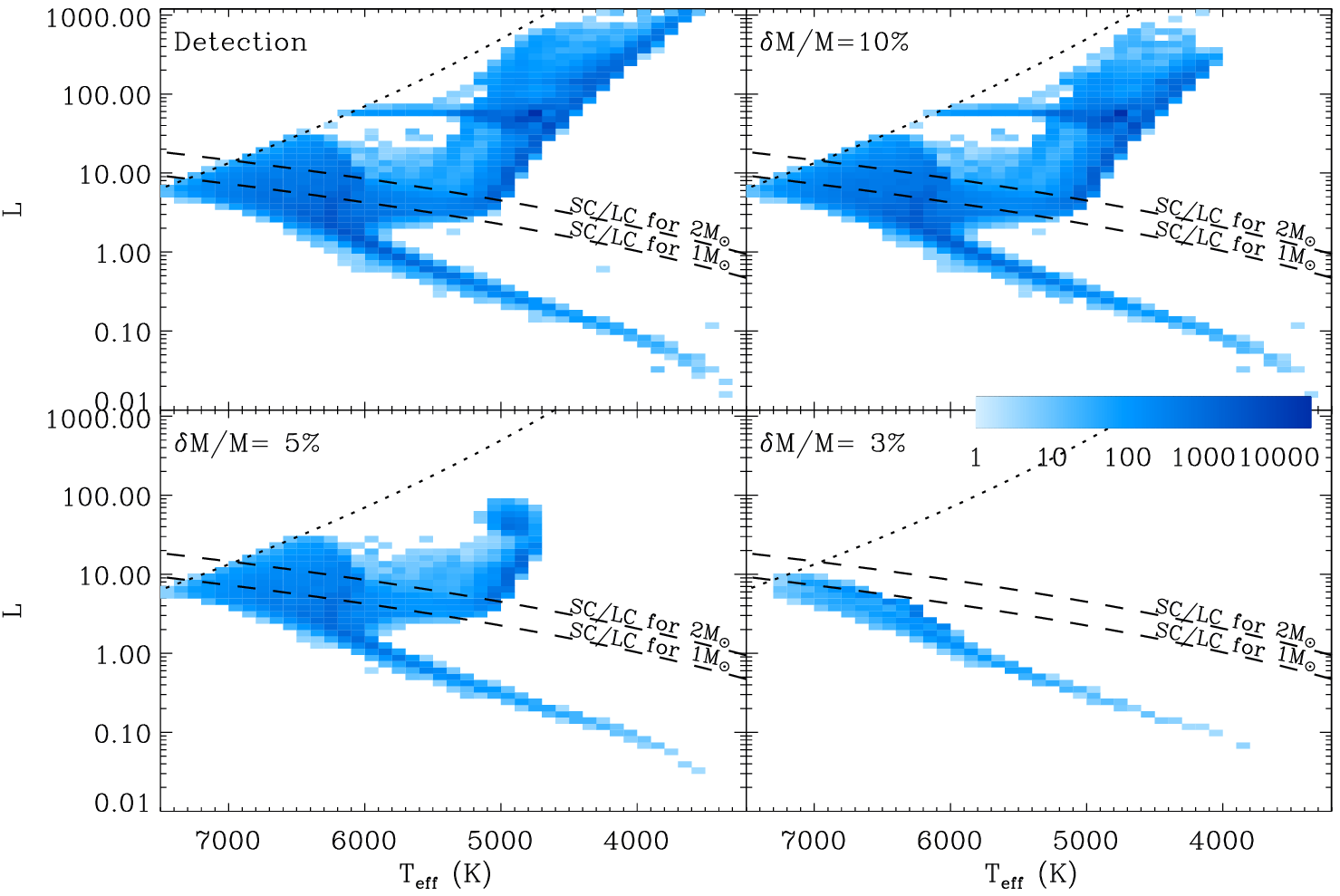}
 \caption{HR diagram of the stars of  Fig. \ref{fig-simu-champ-plato} observable in different conditions.
 The domain where stellar ages can be estimated at the {10\,\%}-level after grid-based modeling corresponds to the condition $\delta M / M=5$\,\% obtained from scaling relations.
  The color denotes the number of stars in bins with $\Delta\Teff = 100$\,K and $\Delta L / L = 20$\,\%. The dotted line shows the red edge of the instability strip; the dashed lines show the limit between the short- and long-cadence acquisition mode of \Plato, for 1 or 2\,$M_\odot$.
 }\label{fig-simu-HR-plato}
\end{figure*}

%---------------------------------------------------------------------------
\section{Discussion\label{discussion}}

\subsection{Spread in the simulation\label{spread-raisons}}

As estimated in Sect. \ref{newcalibration}, the \Kepler\ observations shows a dispersion of about 21\,\% compared to the simulation. Such a scatter is in line with the spread in oscillation amplitudes reported by previous work \citep[e.g.,][]{2011ApJ...743..143H}. We may therefore consider that the spread is related to the diversity of stars and of observation configurations.

An intrinsic source of variability of the seismic signal is related to photometric variability, as noticed in CoRoT and \Kepler\ observations \citep{2009A&A...506...33M,2011ApJ...732L...5C}. \cite{2011ApJ...743..143H} noted a strong correlation between the amplitude signal and the activity directly inferred from the photometric light curve. Metallicity may also play a role through the mode excitation rates, as predicted by \cite{2010A&A...509A..15S}. Its influence on seismic global parameters remains limited; observations of red giants in clusters by \cite{2017A&A...605A...3C} have, however, shown a slight dependence on metallicity of the granulation, and hence of convection. External sources like binarity must be considered at least for two different reasons. First, close binarity is recognized as an efficient cause of damping, as proved by \cite{2014ApJ...785....5G}. Second, binarity dilutes the seismic signal of the oscillatory component \citep{2014ApJ...784L...3M}. Any background source contributing to the photometric signal plays a similar role.

We also tested the seismic variability due to depressed dipole modes in evolved stars
\citep{2012A&A...537A..30M,2014A&A...563A..84G,2017A&A...598A..62M}. These stars show lower EACF coefficients than stars with comparable $\numax$ but regular dipole mode visibilities; the deficit is about 22.5\,\%. They also show lower height-to-background ratios $\HBR$, so that they still follow the generic expression (Eq.~\ref{eqt-EACF-norm}). As a result, even if they participate in the spread observed in
Fig.~\ref{fig-calibration-kepler}, they do not participate in the scatter between observation and simulation.

To sum up, the precision of the predictions derived from our approach remains limited by the variability of the signal with a non-seismic origin. We note that the spread is small and does not hamper the predictive power of our approach, especially for statistical analyses, as performed in the next section. This quality certainly derives from the use of the EACF for detecting the oscillations, which relies on both a strong physical basis and a clear statistical basis. As shown by Fig.~10 of \cite{2018ApJS..239...32P}, the method shows a low fractional dispersion among five different methods. %The low spread can be illustrated in the time domain: a $\pm$20\,\% change of the performance index is compensated by a $\mp$20\,\% change in the observation duration.

\subsection{Core-helium-burning stars\label{evolution}}

Confusion between stars burning either hydrogen in a shell or helium in the core may occur in the $\Dnu$ range from 3 to 8.8\,$\mu$Hz \citep{2014A&A...572L...5M}. Disentangling this confusion is a large success of asteroseismology \citep{2011Natur.471..608B}.
Compared to stars on the red giant branch (RGB), red clump (RC) stars show lower values of $\EACF$, diminished by a factor 2.5. As noted by \cite{2012A&A...537A..30M}, the ratio $\Hmax / \Bmax$ is lower in the red clump. Owing to the definition of $\Hmax$, this is partly, if not totally, due to the fact that RC radial modes have shorter lifetimes \citep{2018A&A...616A..94V}. Here, we do not explore further the physical reasons for the difference but investigate whether the different calibrations (Eqs.~\ref{eqt-EACF-norm} and \ref{eqt-EACF-norm-RC}) can be used to remove the confusion between the evolutionary stages. Therefore, we have analyzed the ratio $\rapEACF$ between the observed $\EACF$ and its value expected on the RGB. The distribution of this ratio shows two clearly separated regions since the difference is significantly larger than the spread of the values (Fig.~\ref{fig-spread}).

The limit value $\rapEACF\ind{lim}=$0.70 allows us to distinguish RC stars from RGB stars. We can use the stars identified as RGB or RC stars from their asymptotic period spacing
\citep{2016A&A...588A..87V} to estimate the possible confusion. In the frequency range [3.3, 8.8\,$\mu$Hz] where confusion is possible \citep{2014A&A...572L...5M}, less than 5\,\% of
the stars with $\rapEACF > \rapEACF\ind{lim}$ are RC stars, whereas less than 9\,\% of the stars with $\rapEACF< \rapEACF\ind{lim}$ are RGB stars. Since a vast majority of the red giants were observed continuously during the whole \Kepler\ mission, such a performance  benefits from the four-year observation duration. We therefore managed to test how the method can discriminate between the different evolutionary stages with shorter data sets, and noted that it requires $\EACF \ge 60$. According to Eq.~\ref{eqt-perf-bf}, this threshold corresponds to an observation duration longer than six months on the RGB or 14 months in the RC. Therefore, the ability to distinguish the different evolutionary stages is not as efficient as determination with the mixed modes. However, the discrimination of the evolutionary stage using $\EACF$ is independent of the determination of the mixed modes, which can be an advantage, for instance, when mixed modes are depressed \citep{2012A&A...537A..30M,2014A&A...563A..84G,2017A&A...598A..62M}.

We note that the ratio $\rapEACF$ keeps its two-component distribution at low frequency, in the domain corresponding to evolved RGB or AGB stars (asymptotic giant branch). We may expect the study of $\rapEACF$ to help us to distinguish AGB from RGB stars, if we assume that the structural properties  that explain low values of $\rapEACF$ in RC stars hold in AGB stars. Three significant features make us confident that this is the case: stars identified as AGB by \cite{2014A&A...572L...5M} according to their period spacings have a low $\rapEACF$; stars identified on the AGB with our new method are hotter than the ones identified on the RGB; and more than 3/4 of the stars with a mass below 0.9\,$M_\odot$ are identified on the  AGB (Fig.~\ref{fig-HR}). The last two tests made use of the APOKASC data \citep{2017AJ....154...94M}. A fourth test, performed with the method proposed by \cite{2012A&A...541A..51K}, provides  similar results (Kallinger, private communication). A dedicated study of the stars identified on the AGB can now be performed, but is beyond the scope of this paper.

\subsection{Clusters\label{clusters}}

We have gathered information from the open clusters observed by \Kepler\ and K2 to draw seismic isochrones in terms of seismic performance \citep[e.g.][]{2011ApJ...729L..10B,2012MNRAS.419.2077M,2016MNRAS.461..760M}. Figure \ref{fig-simu-plato-cluster} shows the performance we can expect for open clusters, depending on their distance modulus $\mu$. Oscillations in main sequence stars are detectable for the closest open clusters only. For instance, the Sun is just above the limit of detectability in an open cluster with $\mu=9$ after two years of observation with \Plato. Similar results cannot be reached with TESS one-month observations, since it would require an open cluster closer than $\mu=2.5$ (30 pc). We did not explore the case of one year of continuous observation with TESS, since such a duration is ensured for stars lying at most 12$^\circ$ away from the ecliptical poles. In practice, results on main sequence stars in clusters are expected with TESS only in the Hyades and Coma Berenices clusters. Many more dimmer clusters with a distance modulus up to ten can be observed with \Plato, with a wide range of ages.

\begin{figure*}
 \sidecaption
 \includegraphics[width=12cm]{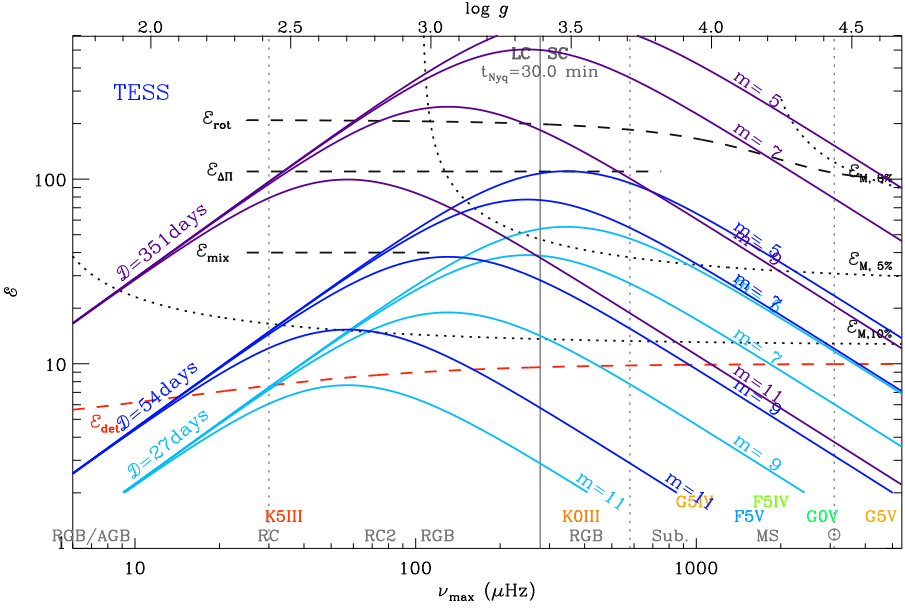}
 \caption{Variation with $\numax$ of the performance index $\EACF$ for TESS, for three different observation durations and two stellar magnitudes. We use the same indications as in Figs.~\ref{fig-calibration-kepler} and \ref{fig-simu-plato-duree}.
 }\label{fig-simu-TESS}
\end{figure*}

\subsection{Typical \Plato\ field of view\label{simuplato}}

The simulation was run for one of the possible fields of the \Plato\ mission, namely a 2230 deg$^2$ region centered at $(l; b)$ = (315$^\circ$; $+30^\circ$), with \textsc{trilegal} \citep{2005A&A...436..895G,2012ASSP...26..165G}. Details of the simulation were given in \cite{2017AN....338..644M}. The simulation considers the following ingredients:\\
 - raw seismic index estimated from Eqs.~\ref{eqt-EACF-norm} or \ref{eqt-EACF-norm-RC};\\
 - spread of the index of \precision\,\% ;\\
 - damping of the index close to the instability strip.\\
Out of the $1.54\times 10^5$ stars brighter than $m\ind{V}=15$ in the field (Fig.~\ref{fig-simu-champ-plato}), $6.9\times 10^4$ show a detectable seismic signal, among which $2.5\times 10^4$ (42\,\%) have a seismic signature high enough to ensure, with seismic scaling relation, an estimate of the mass more precise than 5\,\%. For these stars, one expects the determination of the age to be more precise than 10\,\% with detailed stellar modeling.

The simulation was used to identify which stars in an HR diagram have a seismic signal above a given threshold (Fig.~\ref{fig-simu-HR-plato}). The bottom left panel of Fig.~\ref{fig-simu-HR-plato} shows the stars for which the scaling relations provide an estimate of the mass with a relative uncertainty of 5\,\%. As shown above, this threshold corresponds to stars for which one expects masses with a relative uncertainty of about 3\,\% derived from stellar modeling, hence an estimate of the age of about 10\,\%. We also note that the Nyquist frequency of the long-cadence mode of \Plato\ is high enough to ensure the observation of solar-like oscillations in subgiants, in very good condition with high $\EACF$ factors. The region in the HR diagram accessible with long-cadence measurements only depends on the stellar mass through the seismic proxy $M\,\Teff^{3.5}\ \numax^{-1}$ for $L$.

\subsection{Observations with TESS\label{simutess}}

We used the simulation tool to predict the seismic yield of the NASA/TESS mission \citep{2015JATIS...1a4003R}. We considered the different observation durations defined by the observation procedure, varying from about a month for stars close to the celestial equator up to about a year close to the poles. The simulation was run assuming a noise level floor $\Bsys$ of 60\,ppm\,hr$^{1/2}$ \citep{2016ApJ...830..138C}. As predicted by \cite{2018ApJS..239...34B} and verified in Fig. \ref{fig-simu-TESS}, TESS will excel in observing bright subgiants and early red giants observed in short-cadence mode. Should the noise level be of 60\,ppm\,hr$^{1/2}$, then results for main sequence stars would be modified by a mean shift in magnitude of about $-1.5$ in the instrumental regime.
%---------------------------------------------------------------------------
\section{Conclusion\label{conclusion}}

We propose an analytical method for defining and quantifying the metrics of seismic performance, calibrated with CoRoT and \Kepler\ data. The performance index requires neither  the computation of the synthetic time series, nor the computation of synthetic oscillation spectra. It is derived from scaling relations, in a limited number of steps, each one summarized by a simple analytical expression (Table \ref{tab-simu}): stellar parameters, basic information about the instrumental noise, and the observation duration are enough to derive the  performance index. Then, different thresholds are used to define the detection threshold of the oscillations, to infer information on the stellar core (nuclear regime, rotation rate), or to derive the stellar mass with a given precision.
The spread in the performance is limited, with a standard deviation less than {\precision\,\%} between the simulation and observations, due to stellar variability and binarity.

The method is convenient to analyse the seismic information about a sample of stars located within a given field of view, to compare different conditions of observation, or to compare the seismic output of two fields of view. The simulation of a typical field of view to be observed by the ESA \Plato\ mission shows 10$^5$ stars with detectable oscillations (more than 60000 in the main sequence), among which more than 70\,\% have a seismic result precise enough to predict stellar ages with a relative precision better than 10\,\%. Interestingly, the method seems to provide an efficient way to discriminate between red giants on the RGB and the AGB.

%---------------------------------------------------------------------------
\begin{acknowledgements}

BM, EM, RS, and MJG acknowledge financial support from the Programme National de Physique Stellaire (CNRS/INSU) and from
the SPACEInn FP7 project (SPACEInn.eu). AM, BM, GRD, and LG gratefully acknowledge the support of the International Space Institute (ISSI) for the program AsteroSTEP (Asteroseismology of STEllar Populations). GRD acknowledges the support of the UK Science and Technology Facilities Council.

\end{acknowledgements}
%.....................................................................
\bibliographystyle{aa} % style aa.bst
%\bibliography{biblio_simu}

%---------------------------------------------------------------------------
\begin{appendix}
\section{Classical versus seismic stellar parameters}

\begin{table*}
\tiny \caption{Stellar parameters.\label{tab-valeurs}}
\begin{tabular}{llllllllll}
    \hline
 Star               &Type&$\Teff$ &$\logg$&  $\Dnu$ &$\numax$ & $M$ & $R$ & Ref.\\
                    &    &   (K)&(cm\,s$^{-2}$)& ($\mu$Hz)&($\mu$Hz)& $(M_\odot)$& $(R_\odot)$& \\
\hline
  \multicolumn{8}{c}{Main sequence stars}\\ %%%%%%%%%%%%%%%%%%%%%%%%%%%%%%%%%%%%%%%%%%%%%%%%
\hline
 \object{HD 181420} & F2 & 6580 & 4.17 & 74.9 & 1573 & 1.58 & 1.69 &           2009Bar, 2013Oze \\
 \object{HD 49933 } & F5 & 6780 & 4.24 & 85.8 & 1805 & 1.30 & 1.41 &           2009Ben, 2012Ree \\
 \object{HD 176693} & F8 & 6060 & 4.29 &102.6 & 2151 & 1.03 & 1.20 &  2012App, 2012Mat, 2012Bru \\
 \object{HD 52265 } & G0 & 6100 & 4.29 & 98.5 & 2150 & 1.27 & 1.34 &           2011Bal, 2013Giz, 2014Leb \\
 \object{16 Cyg A } &G1.5& 5825 & 4.29 &102.5 & 2180 & 1.11 & 1.24 &                    2012Met \\
 \object{\aCenA   } & G2 & 5790 & 4.33 &105.7 & 2410 & 1.10 & 1.23 &          2004Bed, 2002The  \\
       Sun          & G2 & 5777 & 4.43 &134.4 & 3050 & 1.00 & 1.00 &                        Sun \\
 \object{16 Cyg B } &G2.5& 5750 & 4.36 &115.3 & 2578 & 1.07 & 1.13 &                    2012Met \\
      KIC 176465 A  & G4 & 5830 & 4.46 &146.8 & 3260 & 0.96 & 0.90 &  2017Whi\\
      KIC 176465 B  & G4 & 5740 & 4.49 &155.4 & 3520 & 0.93 & 0.89 &  2017Whi\\
 \object{HD 173701} & G8 & 5390 & 4.47 &148.5 & 3444 & 1.00 & 0.93 &  2012Mat, 2012App, 2012Bru \\
 \object{HD 188753 B}&G9 & 5500 & 4.46 &147.4 & 3274 & 0.86 & 0.91 &  2018Mar \\
\object{\aCenB   }  & K1 & 5260 & 4.54 &161.4 & 4090 & 0.91 & 0.86 &          2005Kje, 2002The  \\
\hline
  \multicolumn{8}{c}{Subgiants} \\%%%%%%%%%%%%%%%%%%%%%%%%%%%%%%%%%%%%%%%%%%%%%%%%%%%%%%%%%%%%
\hline
 \object{Procyon  } & F5 & 6550 & 3.94 & 54.1 &  918 & 1.46 & 2.04 &           2010Bed, 2008Mos \\
 \object{HD 185834} & F6 & 6020 & 3.95 & 55.1 &  988 & 1.17 & 1.91 &           2012App, 2012Bru \\
      KIC  7976303  & F8 & 6260 & 3.88 & 51.1 &  833 & 1.17 & 2.03 &           2012App, 2012Mat \\
      KIC 10273246  & F9 & 6150 & 3.92 & 48.8 &  913 & 1.37 & 2.19 &           2011Cam, 2012Cre \\
 \object{HD 49385 } & G0 & 6095 & 3.95 & 55.5 &  994 & 1.25 & 1.94 &           2010Deh, 2011Deh \\
      KIC 10920273  &G1-2& 5880 & 3.96 & 57.2 & 1026 & 1.23 & 1.88 &           2011Cam, 2012Cre \\
      KIC 7107778 A & G8 & 5180 & 3.65 & 32.1 &  544 & 1.42 & 2.93 & 2018LiY  \\
      KIC 7107778 B & G8 & 5180 & 3.70 & 34.7 &  594 & 1.39 & 2.76 & 2018LiY  \\
 \object{HD 182736} & G8?& 5261 & 3.68 & 34.6 &  568 & 1.30 & 2.70 &                    2012Hub \\
\hline
  \multicolumn{8}{c}{RGB} \\%%%%%%%%%%%%%%%%%%%%%%%%%%%%%%%%%%%%%%%%%%%%%%%%%%%%%%%%%%%%%%%%
\hline
\object{HD 181907}  & K1 & 4790 & 2.35 & 3.41 & 28.0 & 1.20 &12.20 &          2010Car, 2010Mig  \\
      KIC 4044238   & K5 & 4800 & 2.43 & 4.07 & 33.7 &      &      &                    2012Mos \\
\object{HD 50890 }  & K2 & 4670 & 2.08 & 1.71 & 15.0 & 4.20 &29.90 &                    2012Bau \\
      KIC 11618103  & G5 & 4870 & 2.97 & 9.38 &  115 & 1.45 & 6.60 &                    2011Jia \\
      KIC 4072740   & K1 & 4882 & 3.33 & 18.4 & 261  &      &      &               2013Mos, Mol2013\\
      KIC 7341231   & G2 & 5300 & 3.53 & 28.8 &  399 & 0.83 & 2.62 &                  2012Deh   \\
\hline
  \multicolumn{8}{c}{RC and secondary RC}\\ %%%%%%%%%%%%%%%%%%%%%%%%%%%%%%%%%%%%%%%%%%%%%%
\hline
      KIC 9655101   & G8 & 5039 & 2.90 & 7.82 & 97.2 & 2.31 & 8.86 & Mol2013, Mos2013\\
      KIC 9716522   & G9 & 4952 & 2.64 & 4.84 & 53.8 & 2.53 & 12.6 & Mol2013, Mos2013\\
\hline
\end{tabular}

\textbf{References.} \\ %%%% IL FAUDRA SUPPRIMER LES \\ ET LES REMPLACER PAR DES ;
2002The = \cite{2002A&A...392L...9T}\suivant
2004Bed = \cite{2004ApJ...614..380B}\suivant
2005Kje = \cite{2005ApJ...635.1281K}\suivant
2008Mos = \cite{2008A&A...478..197M}\suivant
2009Bar = \cite{2009A&A...506...51B}\suivant
2009Ben = \cite{2009A&A...507L..13B}\suivant
2010Bed = \cite{2010ApJ...713..935B}\suivant
2010Car = \cite{2010A&A...509A..73C}\suivant
2010Deh = \cite{2010A&A...515A..87D}\suivant
2010Mig = \cite{2010A&A...520L...6M}\suivant
2011Bal = \cite{2011A&A...530A..97B}\suivant
2011Cam = \cite{2011A&A...534A...6C}\suivant
2011Deh = \cite{2011A&A...535A..91D}\suivant
2011Jia = \cite{2011ApJ...742..120J}\suivant
2012App = \cite{2012A&A...543A..54A}\suivant
2012Bau = \cite{2012A&A...538A..73B}\suivant
2012Bru = \cite{2012MNRAS.423..122B}\suivant
2012Cre = \cite{2012A&A...537A.111C}\suivant
2012Hub = \cite{2012ApJ...760...32H}\suivant
2012Mat = \cite{2012ApJ...749..152M}\suivant
2012Met = \cite{2012ApJ...748L..10M}\suivant
2013Oze = \cite{2013A&A...558A..79O}\suivant
2012Ree = \cite{2012A&A...539A..63R}\suivant
2013Giz = \cite{2013PNAS..11013267G}\suivant
2013Mol = \cite{2013MNRAS.434.1422M}\suivant
2013Mos = \cite{2013A&A...550A.126M}\suivant
2014Leb = \cite{2014A&A...569A..21L}\suivant
2017Whi = \cite{2017A&A...601A..82W}\suivant
2018Mar = \cite{2018A&A...617A...2M}\suivant
2018LiY = \cite{2018MNRAS.476..470L}\suivant
\end{table*}

In order to link the classical and seismic parameters, we present in Table \ref{tab-valeurs} a set of stars that benefitted from seismic analysis and modeling. The $\logg$ were estimated by the relation Eq.~\ref{eqt-numax-logg}; the stellar types are given as they appear in the literature \citep[most often][for \Kepler\ stars]{2013MNRAS.434.1422M}.

\end{appendix}

\end{document}